\documentclass[twocolumn,aps,pre,epsf,superscriptaddress]{revtex4-2}
\usepackage{amsmath}
\usepackage{epsfig}
\usepackage{color}
\usepackage{epstopdf}
\usepackage{amssymb}
\usepackage{multirow}
\usepackage{hyperref}
\usepackage[greek,ukrainian,english]{babel}
\usepackage[T2A]{fontenc}
\usepackage{teubner}
\usepackage{psfrag}

\begin{document}

\title{Finite-size scaling of the random-field Ising model above the upper critical dimension}

\author{Nikolaos~G.~Fytas}
\affiliation{Department of Mathematical Sciences, University of Essex, Colchester CO4 3SQ, United Kingdom}

\author{V\'{i}ctor Mart\'{i}n-Mayor}
\affiliation{Departamento de F\'isica
  T\'eorica I, Universidad Complutense, 28040 Madrid, Spain}
\affiliation{Instituto de Biocomputac\'ion y F\'isica de Sistemas
Complejos
  (BIFI), 50009 Zaragoza, Spain}

\author{Giorgio Parisi}
\affiliation{Dipartimento  di  Fisica,  Sapienza Universit\`{a} di
Roma,  P.le  Aldo  Moro  2, 00185  Rome, Italy and INFN, Sezione
di Roma I,  IPCF -- CNR,  P.le  A. Moro 2, 00185 Rome, Italy}

\author{Marco Picco}
\affiliation{Laboratoire de Physique Th\'{e}orique et Hautes Energies, UMR7589, Sorbonne Universit\'{e} et CNRS, 4 Place Jussieu,
75252 Paris Cedex 05, France}

\author{Nicolas Sourlas}
\affiliation{Laboratoire de Physique Th\'eorique de l'Ecole
Normale Sup\'erieure (Unit{\'e} Mixte de Recherche du CNRS et de l'Ecole Normale
  Sup\'erieure, associ\'ee \`a l'Universit\'e Pierre et Marie Curie, PARIS VI)
  24 rue Lhomond, 75231 Paris Cedex 05, France}

\date{\today}

\begin{abstract}
Finite-size scaling above the upper critical dimension is a long-standing puzzle in the field of Statistical Physics. Even for pure systems various scaling theories have been suggested, partially corroborated by numerical simulations. In the present manuscript we address this problem in the even more complicated case of disordered systems. In particular, we investigate the scaling behavior of the random-field Ising model at dimension $D = 7$, \emph{i.e.}, above its upper critical dimension $D_{\rm u} = 6$, by employing extensive ground-state numerical simulations. Our results confirm the hypothesis that at dimensions $D > D_{\rm u}$, linear length scale $L$ should be replaced in finite-size scaling expressions by the effective scale $L_{\rm eff} = L^{D / D_{\rm u}}$. Via a fitted version of the quotients method that takes this modification, but also subleading scaling corrections into account, we compute the critical point of the transition for Gaussian random fields and provide estimates for the full set of critical exponents. Thus, our analysis indicates that this modified version of finite-size scaling is successful also in the context of the random-field problem.
\end{abstract}

\maketitle

\section{Introduction}
\label{sec:introduction}

The random-field Ising model (RFIM) represents one of the simplest models of cooperative behavior with quenched disorder~\cite{imry:75,aharony:76,young:77,fishman:79,parisi:79,cardy:84,imbrie:84,villain:84,bray:85b,fisher:86b,schwartz:85,gofman:93,rieger:95,belanger:98,barber:01,hartmann:01}.Despite being seemingly simple in terms of definition, the combined presence of random fields and the standard Ising behavior accounts for a vast range of new physical phenomena, many of them remain unresolved even after 50 years of extensive research. Additionally, its direct relevance to two- and three-dimensionsal experimental analogues in condensed-matter physics, such as diluted antiferromagnets in a field, colloid-polymer mixtures, and others~\cite{belanger:83,belanger:91,belanger:98,vink:06} establishes the RFIM as one of the most prominent platform models for the designing and/or deciphering of experiments. Another asset is that a vast majority of non-equilibrium phenomena including critical hysteresis, avalanches, and the Barkhausen noise~\cite{sethna:93,perkovic:99,sethna:06,shukla:18}  can be studied through the RFIM. For a review but also a summary of most recent results we refer to Ref.~\cite{rychkov:23}. 

It is well established that the physically
relevant dimensions of the RFIM lay between $2 < D < 6$, where
$D_{\rm l} = 2$ and $D_{\rm u} = 6$ are the lower and upper
critical dimensions of the model, respectively~\cite{imry:75}. 
Although the critical behavior of the RFIM at these dimensions has been scrutinized by a variety of methods, 
a consensus has not been reached for decades. Fortunately, over the last few years several
ambiguities have been put at ease
due to the development of a powerful panoply of
simulation and statistical analysis methods, that have
set the basis for a fresh revision of the problem~\cite{fytas:16b}. 
In fact, some
of the main controversies have been resolved, the most notable
being the illustration of critical universality in terms of
different random-field distributions~\cite{fytas:13,fytas:16,fytas:17} -- see also Ref.~\cite{picco:15} where it was shown that the diluted Ising
model in a field belongs also to the same universality class with the RFIM as
predicted by the perturbative renormalization group -- and the restoration
of supersymmetry and dimensional reduction at $D =
5$~\cite{fytas:17,fytas:17b,fytas:18,fytas:19,fytas:19b}. We refer the reader to
Refs.~\cite{tissier:11,tissier:12,tarjus:13,hikami:18,kaviraj:21,kaviraj:21b} for
additional evidence supporting this latter respect.  
Furthermore the large-scale numerical simulations
of Refs.~\cite{fytas:13,fytas:16,fytas:17,fytas:17b,fytas:19b} have provided
high-accuracy estimates for the full spectrum of critical exponents, putting at rest previous fears of possible 
violations of fundamental scaling relations.

On the other hand for $D \geq D_{\rm u}$ the RFIM is expected to show dimension-independent 
mean-field behavior~\cite{imry:75}, with the critical exponents holding the mean-field values of the pure Ising ferromagnet (exactly at $D = D_{\rm u}$ the well-known logarithmic corrections appear~\cite{kenna:91,kenna:04,ahrens:11}). 
At this point we should emphasize that although the method of finite-size scaling has been successfully applied to the analysis of results by numerous numerical simulations for spin models at $D < D_{\rm u}$, the situation becomes more complicated when one considers the system above its $D_{\rm u}$, as discussed extensively for the 5D Ising model (note that $D_{\rm u}=4$ for the pure Ising ferromagnet)~\cite{mon:96,luijten:96,parisi:96b,luijten:97,luijten:99,jones:05,berche:12,kenna:13,wittmann:14b,berche:22}. 

For periodic boundary conditions a possible solution has been proposed. The key point in these studies~\cite{mon:96,luijten:96,parisi:96b,luijten:97,luijten:99,jones:05,berche:12,kenna:13,wittmann:14b,berche:22} is that at dimensions $D > D_{\rm u}$ the linear length scale $L$ of the system should be replaced in finite-size scaling expressions by a new effective length scale of the form $L_{\rm eff} = L^{D/D_{\rm u}}$, an ansatz originally proposed by Kenna and Lang in the framework of the $\phi^{4}$  theory~\cite{kenna:91}. In fact, the ratio $D / D_{\rm u}$ is the so-called \textgreek{\coppa} exponent introduced by Kenna and Berche in Ref.~\cite{kenna:13} and elaborated in several subsequent works -- see Ref.~\cite{berche:22} where an overview of the renormalization group as a successful framework to understand critical phenomena above the upper critical dimension is provided. The proposed scaling theory not only concerns the Ising model but it is believed to be more general. In particular, the finite-size scaling of percolation above its upper critical dimension has also been successfully analyzed in the same framework~\cite{kenna:17}. However, we should point out that the problem is highly non-trivial as the selection of boundary conditions qualitatively changes the scaling~\cite{sola:16}, so that the case of free (or other type of) boundary conditions is not yet completely settled~\cite{berche:12,berche:22}.

What is even more, for disordered systems, and in particular for the RFIM, not much has been achieved in this direction, with the exception of Ref.~\cite{ahrens:11} where a qualitative picture of the transition has been provided at high dimensions. In the context of spin glasses, see Ref.~\cite{Aspelmeier_2008}. To this end, we present in the current work an extensive numerical study of the RFIM at $D = 7$ using exact ground-state simulations and a suitable finite-size scaling method based on phenomenological renormalization that takes into account the effective length scale $L_{\rm eff}$. We locate the critical point of the transition for Gaussian fields and monitor the size evolution of effective critical exponents. Our final results are compatible up to a very good numerical accuracy with their mean-field expectations. Instrumental in our analysis is the use of a proper value for the corrections-to-scaling exponent $\omega$. In this respect, we provide in Appendix~\ref{appendix} a detailed derivation of $\omega$ for the large-$N$ limit of the $O(N)$ model, starting from Br\'ezin’s analysis~\cite{brezin:82}. We find that the exponent $\omega$ corresponding to the $O(N)$ model plays a crucial role for a safe determination of the critical properties in the 7D RFIM.

The remainder of this manuscript is as follows: In Sec.~\ref{sec:model} the model and methods employed are described shortly and in Sec.~\ref{sec:results} our main results on the scaling aspects of the 7D RFIM are presented. We conclude in Sec.~\ref{sec:summary} by providing a summary and an outlook for future work in this direction. 

\section{Model and methods}
\label{sec:model}

The Hamiltonian of the RFIM is
\begin{equation}
	\label{H} 
	{\mathcal H} = - J \sum_{\langle xy\rangle} S_x S_y - \sum_{x} h_x S_x,\;
\end{equation}
with the spins $S_x = \pm 1$ on a $D=7$ hypercubic lattice with
periodic boundary conditions and energy units $J=1$, and
$h_x$ independent random magnetic fields with zero mean and
variance $\sigma^{2}$. Given our previous universality confirmations~\cite{fytas:13,fytas:16,fytas:17}, 
we have restricted ourselves to Gaussian normal-distributed $\{h_x\}$. We work directly at zero temperature~\cite{ogielski:86,auriac:85,middleton:01,middleton:02,middleton:02b} because the relevant fixed point of the model lies there~\cite{villain:84,bray:85b,fisher:86b}. The system has a ferromagnetic phase at small $\sigma$, that, upon increasing the disorder, becomes paramagnetic at the critical point $\sigma_{\rm c}$. Obviously, the only relevant spin configurations are ground
states, which are non-degenerate for continuous random-
field distributions. An instance of random
fields $\{h_x\}$ is named a sample and thermal mean values 
are denoted as $\langle \cdots \rangle$. The subsequent average over samples is indicated by an overline, (\emph{e.g}., for the magnetization density $m=\sum_{x}S_x/L^D$, we consider both $\langle m \rangle$ and $\overline{\langle m \rangle}$). 

\begin{figure}
	\includegraphics[width=8.5 cm]{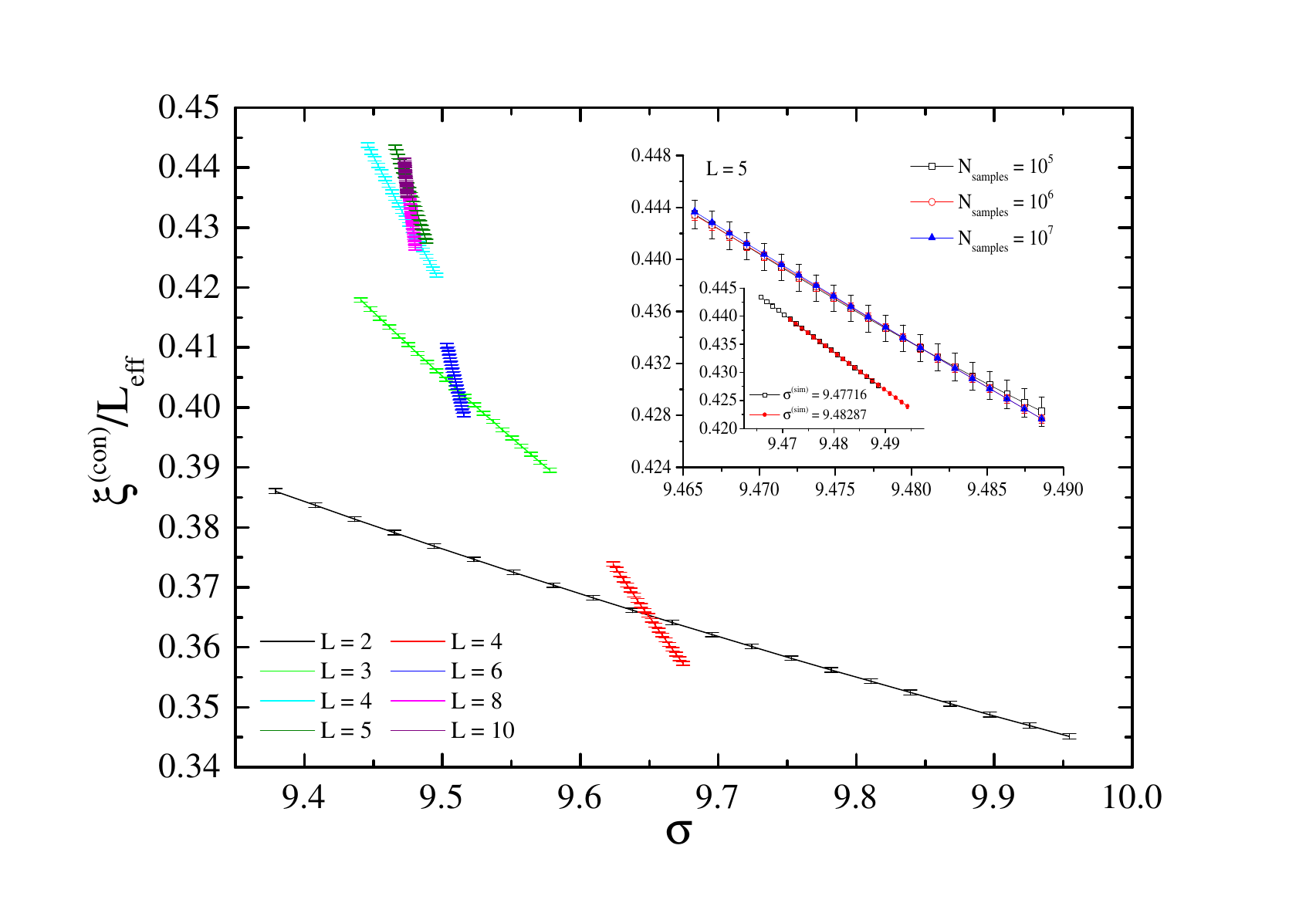}
	\caption{\label{fig:quotients} Connected correlation length in units of the effective system size as a function of the random-field strength $\sigma$. Lines join data obtained from reweighting extrapolation. The inset presents typical illustrations with respect to the sample-to-sample fluctuations and the errors induced by the reweighting extrapolation for a size $L=5$. $N_{\rm samples}$ denotes the number of disorder realizations and $\sigma^{\rm (sim)}$ the value of random field at which the simulation was performed. The comparative data for the different values of $N_{\rm samples}$ refer to the simulation value $\sigma^{\rm (sim)} = 9.47716$ and serve in favor of our numerical scheme.}
\end{figure}

The scaling theory of the RFIM entails an analysis of two correlation functions, namely the
connected and disconnected propagators $C^{\rm (con)}_{xy}$ and
$C^{\rm (dis)}_{xy}$~\cite{bray:85b,fisher:86b}:
\begin{equation}\label{eq:anomalous}
	C^{\rm (con)}_{xy}\equiv\frac{\partial\overline{\langle
			S_x\rangle}}{\partial h_y} \,,\
	C^{\rm (dis)}_{xy}\equiv\
	\overline{\langle S_x\rangle\langle S_y\rangle}\,.
\end{equation}
For each of these two propagators we scrutinize the second-moment
correlation lengths~\cite{amit:05}, denoted as $\xi^{\rm (con)}$ and
$\xi^{\rm (dis)}$, respectively. Hereafter, we shall indicate
with the superscript ``${\rm (con)}$'', \emph{e.g.},
$\xi^{\rm (con)}$, quantities computed from the connected
propagator. Similarly, the superscript ``${\rm (dis)}$'',
\emph{e.g.}, $\xi^{\rm (dis)}$, will refer to the  propagator
$C^{\mathrm{(dis)}}$. We also compute the corresponding
connected susceptibility $\chi^{\rm (con)}$ to obtain 
the anomalous dimension $\eta$, as well the dimensionless Binder ratio
$U_4 = \overline{\langle m ^4 \rangle}/\overline{\langle
	m^2\rangle}^2$.

As it is well-known, the random field is a relevant perturbation at the pure fixed
point, and the random-field fixed point is at
$T=0$~\cite{villain:84,bray:85b,fisher:86b}. The main assumption leading to this result is that the scale of variation of the effective free energy in a correlation volume scales as $\xi^{\theta}$, where $\theta$ is so-called violation of hyperscaling exponent~\cite{fisher:86b}. This is a consequence of the observation that the important competition yielding the phase transition is between the exchange interactions and the random field which implies that the controlling critical fixed point is at zero temperature with the temperature irrelevant. Hence, the critical
behavior is the same everywhere along the phase boundary and we can predict it simply by staying at $T=0$ and crossing the phase boundary at the critical field point. This is a convenient approach because we can determine the ground states of the system
exactly using efficient optimization
algorithms~\cite{ogielski:86,auriac:86,sourlas:99,hartmann:95,auriac:97,swift:97,bastea:98,hartmann:99,hartmann:01,middleton:01,middleton:02,middleton:02b,dukovski:03,wu:05,fytas:13,alava:01,ahrens:11,stevenson:11}
through an existing mapping of the ground state to the
maximum-flow optimization
problem~\cite{auriac:85,cormen:90,papadimitriou:94}. A clear advantage of this approach is the ability to simulate large system sizes and disorder ensembles in rather moderate computational
times. The application of maximum-flow algorithms to the RFIM is nowadays well established~\cite{alava:01}. One of the most efficient network flow algorithms used to solve the RFIM is the push-relabel algorithm of
Tarjan and Goldberg~\cite{goldberg:88}. In
the present study we prepared our own C version of the algorithm
that involves a modification proposed by Middleton \emph{et
al.}~\cite{middleton:01,middleton:02,middleton:02b} that removes
the source and sink nodes, reducing memory usage and also
clarifying the physical
connection~\cite{middleton:02,middleton:02b}. Further details on the numerical implementation can be found in Ref.~\cite{fytas:16b}.

One big advantage of our numerical toolkit is that it allows us from simulations at a given $\sigma$ to compute $\sigma$-derivatives
and extrapolate to neighboring $\sigma$ values by means of a
reweighting method -- see Ref.~\cite{fytas:16b} for
full mathematical derivations of fluctuation-dissipation and reweighting formulas. 
In the present work we consider lattice sizes within the range $L _{\rm min}=2$ to $L_{\rm
max}=10$. For each pair of $\{L, \;\sigma\}$ values we compute exact ground
states for $10^6$ samples (initial exploratory runs were performed using $10^5$ samples), outperforming previous studies -- For comparison, $5000$ samples with  $L_{\rm max} = 8$ were used in Ref.~\cite{ahrens:11}. 

We follow the quotients method for finite-size
scaling~\cite{amit:05,nightingale:76,ballesteros:96}, taking into account the modification
$L \rightarrow L_{\rm eff} = L^{7/6}$, as we work above the upper critical dimension with periodic boundary conditions. As mentioned above, Kenna and Berche identify this ratio $7/6$ as the new critical exponent \textgreek{\coppa}, giving extensive discussions in Refs.~\cite{kenna:13,berche:22}.
In practice, we focus on three dimensionless quantities $g(\sigma,L_{\rm eff})$ that,
barring correction to scaling, are independent of the system size at the critical
point, namely $\xi^{\mathrm{(con)}}/L_{\rm eff}$, $\xi^{\mathrm{(dis)}}/L_{\rm eff}$, and $U_4$.
Given a dimensionless quantity $g$, we consider a pair of lattices
sizes $(L_{\rm eff}, 2L_{\rm eff})$ and determine the crossing
$\sigma_{\mathrm{c},L_{\rm eff}}$, where $g(\sigma_{\mathrm{c},L_{\rm eff}},L_{\rm eff})=
g(\sigma_{\mathrm{c},L_{\rm eff}},2L_{\rm eff})$, see Fig.~\ref{fig:quotients}. This allows us to compute three such
$\sigma_{\mathrm{c},L_{\rm eff}}$, a first for $\xi^{\mathrm{(con)}}/L_{\rm eff}$,
another for $\xi^{\mathrm{(dis)}}/L_{\rm eff}$, and a third for $U_4$.

Dimensionful quantities $O$ scale with $\xi$ in the thermodynamic
limit as $\xi^{x_O/\nu}$, where $x_O$ is the scaling dimension of
$O$ and $\nu$ the critical exponent of the correlation length. At finite system sizes
we consider the quotient $Q_{O,L_{\rm eff}} =
O_{2L_{\rm eff}}/O_{L_{\rm eff}}$ at the crossing
\begin{equation}\label{eq:QO}
	Q_{O,L_{\rm eff}}^\mathrm{cross} = 2^{\frac{7}{6}\frac{x_O}{\nu}} +
	O(L_{\rm eff}^{-\omega}).
\end{equation}
$Q_{O,L_{\rm eff}}^\mathrm{cross}$ can be
evaluated at the crossings of $\xi^{\mathrm{(con)}}/L_{\rm eff}$,
$\xi^{\mathrm{(dis)}}/L_{\rm eff}$, and $U_4$. Renormalization group
tells us that $x_O$, $\nu$, and the leading
corrections-to-scaling exponent  $\omega$ are universal.
Instances of dimensionful quantities used in this work are the derivatives
of correlation lengths $\xi^{\rm (con)}$ and $\xi^{\rm (dis)}$ $[x_{D_\sigma \xi^{\mathrm{(con)}}}=x_{D_\sigma \xi^{\mathrm{(dis)}}}=1+\nu]$ 
and the connected susceptibility $[x_{\chi^{\mathrm{(con)}}}= \nu(2-\eta)]$.
Scaling corrections for the critical point are of order $L_{\rm eff}^{-(\omega+\frac{1}{\nu})}$, $L_{\rm eff}^{-(2\omega+\frac{1}{\nu})}$, etc. Note that as we applied the quotients method at the crossings of $\xi^{\rm
(con)} / L_{\rm eff}$, $\xi^{\rm (dis)} / L_{\rm eff}$, and $U_4$, the data
sets of our simulations were tripled for each pair of system sizes
used and thus our practice was to use joint fits imposing a common extrapolation to the thermodynamic limit. Finally, the exponent $\omega$ is fixed to the value $\omega = 1/2$ throughout the analysis below, see Appendix~\ref{appendix}.

\begin{figure}
	\centering
	\includegraphics[width=8.5 cm]{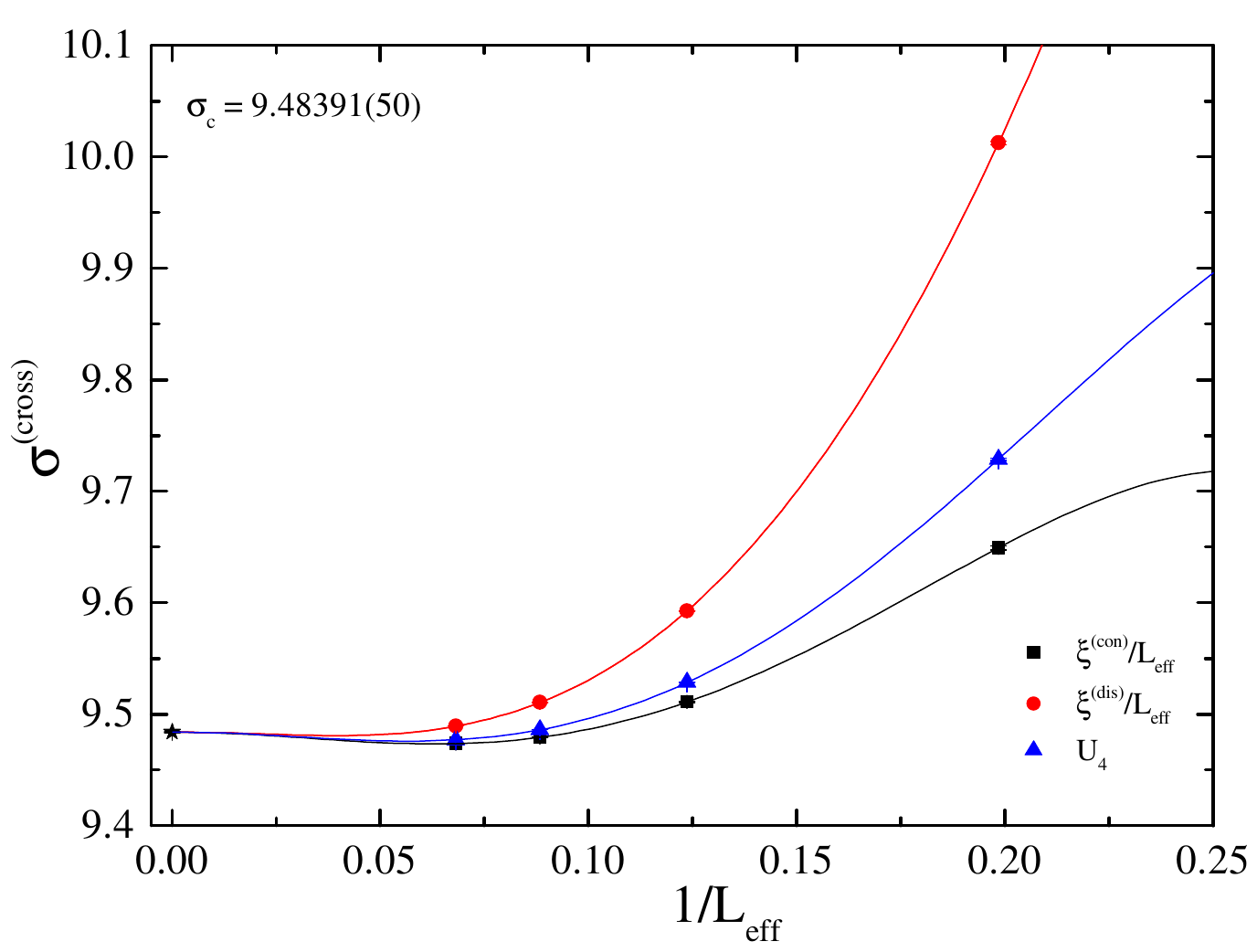}
	\caption{\label{fig:critical_point} Crossing points $\sigma_{\mathrm{c},L_{\rm eff}}$ vs. $1/L_{\rm eff}$.}
\end{figure}

Finally, some comments on the fitting procedure: We restrict ourselves
to data with $L\geq L_{\rm min}$ and to determine an
acceptable $L_{\rm min}$ we employ the standard $\chi^{2}/{\rm DOF}$-test
for goodness of fit, where $\chi^2$ is computed using the
complete covariance matrix and DOF denotes the number of degrees of freedom. Specifically, we consider a fit as being fair only if $10\% < Q < 90\%$, where $Q$ denotes the probability of finding a $\chi^{2}$ value which is even larger than the one actually found from our data~\cite{press:92}.

\begin{figure}
	\centering
	\includegraphics[width=8.5 cm]{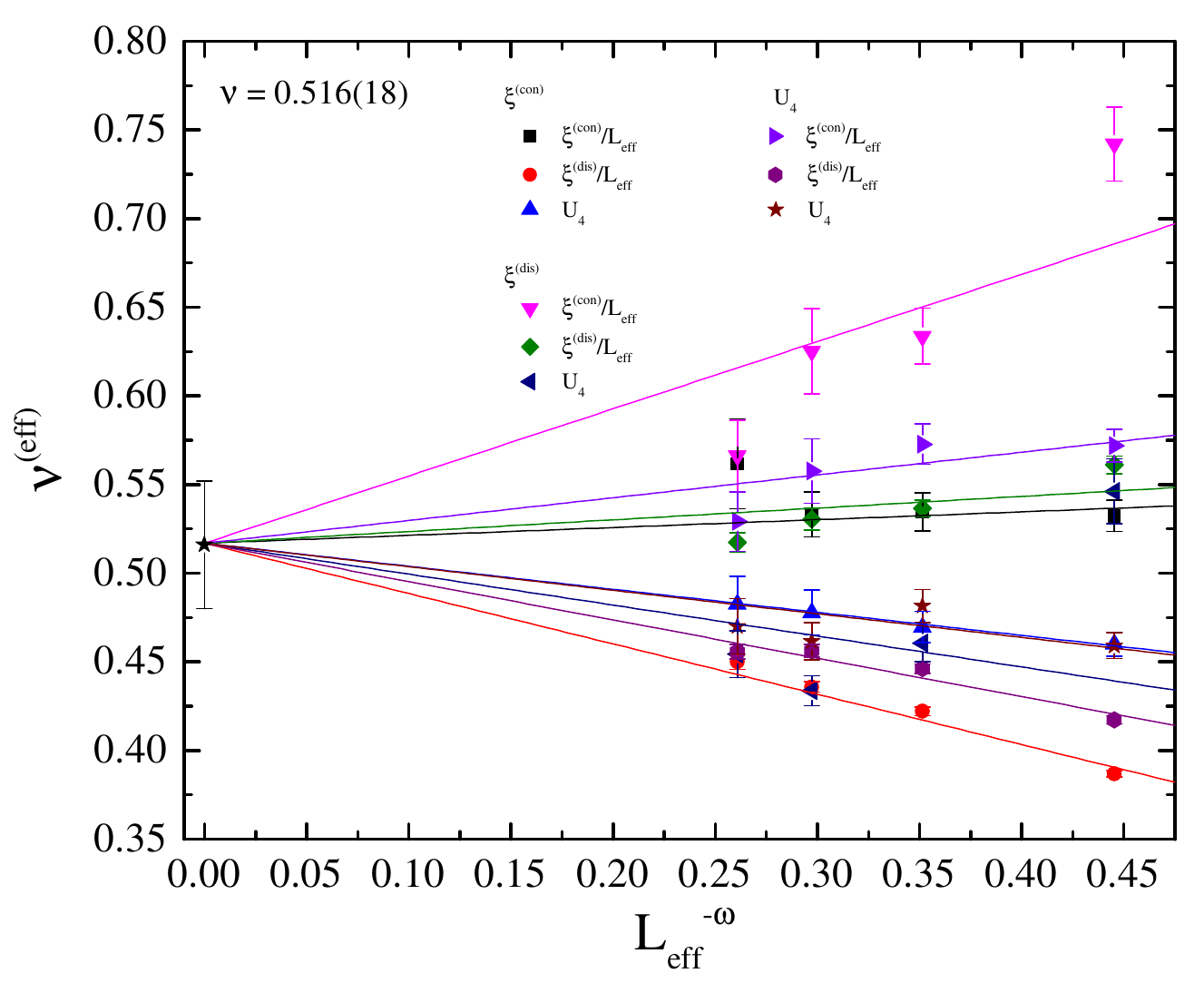}
	\caption{\label{fig:exponent_nu} Effective critical exponent $\nu^{\rm (eff)}$ vs. $L_{\rm eff}^{-\omega}$. Results are shown obtained from the derivatives of $\xi^{\rm (con)}$, $\xi^{\rm (dis)}$, and $U_{4}$ for all data sets at hand.}
\end{figure}

\section{Results}
\label{sec:results}

\begin{figure}
	\centering
	\includegraphics[width=8.5 cm]{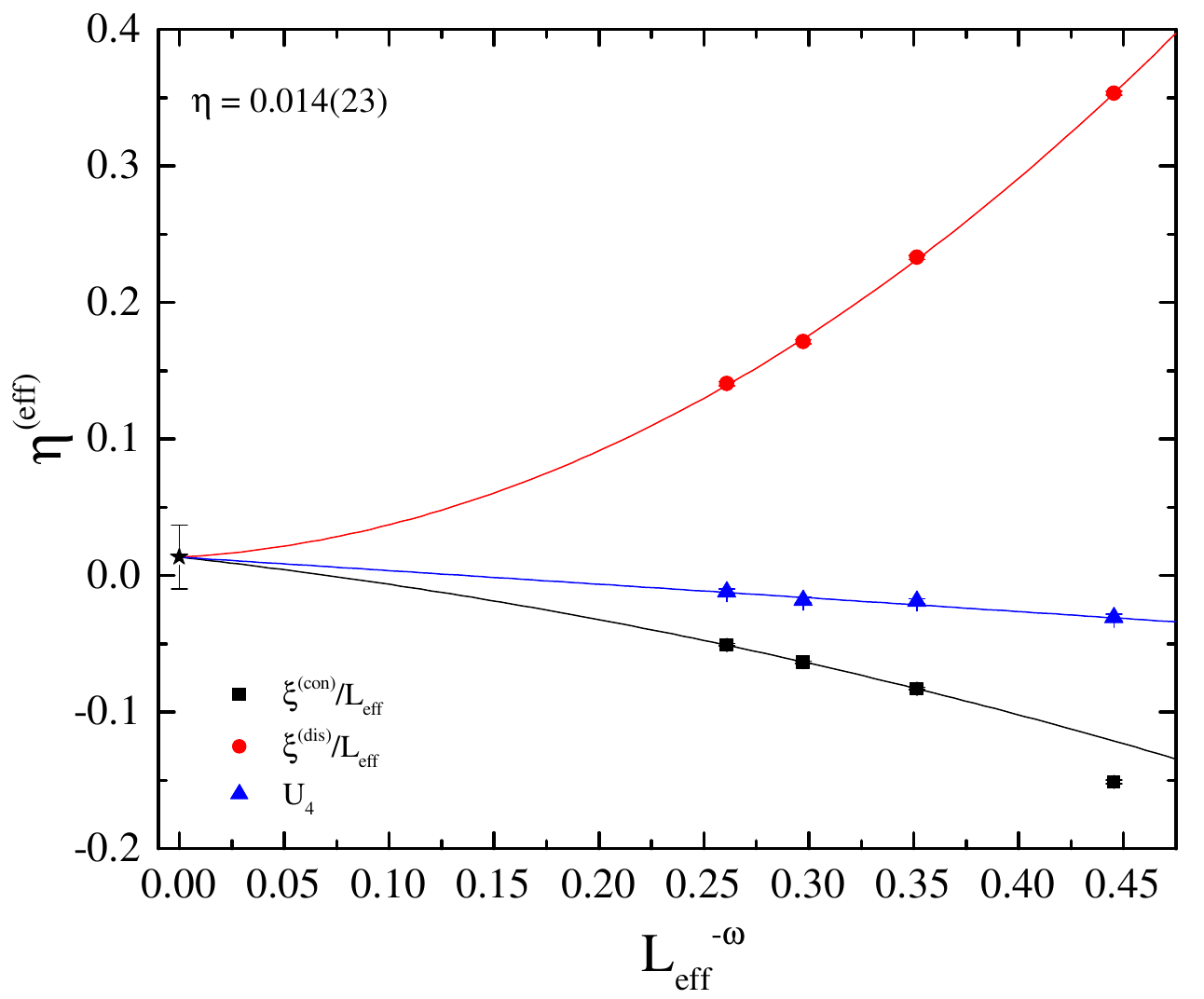}
	\caption{\label{fig:exponent_eta} Effective critical exponent $\eta^{\rm (eff)}$ vs. $L_{\rm eff}^{-\omega}$.}
\end{figure}

We start the presentation of our results in Fig.~\ref{fig:critical_point} where a joint fit of the form
\begin{equation}
	\label{eq:QFS}
	\sigma_{\mathrm{c},L_{\rm eff}}=\sigma_c+ b_1 L_{\rm eff}^{-(\omega+\frac{1}{\nu})}+
	b_2 L_{\rm eff}^{-(2\omega+\frac{1}{\nu})} + b_3 L_{\rm eff}^{-(3\omega+\frac{1}{\nu})}\;,
\end{equation}
provides the estimate $\sigma_{\rm c} = 9.48391(50)$ for the critical field, in excellent agreement (but higher numerical accuracy) with the earlier result $9.48(3)$ of Ref.~\cite{ahrens:11}. The coefficients $b_{k}$ with $k=1,2,3$ are just scaling amplitudes and the quality is quite good ($Q \sim 45\%$). Figures~\ref{fig:exponent_nu} and \ref{fig:exponent_eta} document the infinite-limit size extrapolations of the main critical exponents $\nu$ and $\eta$ using also joint fits of the form~(\ref{eq:QO}) in linear and quadratic $L_{\rm eff}^{-\omega}$ order and with cutoff sizes $L_{\rm min} = 2$ and  $3$, respectively. In both cases a fair fit quality is obtained, namely $Q\sim 25\%$ and $18\%$, respectively. Evidently, the obtained estimates $\nu = 0.516(18)$ and $\eta = 0.014(23)$ are  compatible to the mean-field (MF) values $\nu^{\rm (MF)}=1/2$ and $\eta^{\rm (MF)}=0$.

Obtaining the critical exponent $\alpha$ of the specific heat is much more trickier in most cases, and the random-field problem is no exception~\cite{fytas:16b,fytas:17b,fytas:19b,hartmann:01}. The specific heat of the RFIM can be computed via ground-state calculations and the bond-energy density
$E_{J}$~\cite{holm:97}. This is the first derivative $\partial E/\partial J$ of the ground-state energy with respect to the random-field strength $\sigma$~\cite{middleton:01,hartmann:01}. The $\sigma$-derivative of the sample averaged quantity $\overline{E}_{J}$ then gives the second derivative with respect to $\sigma$ of the total energy and thus the sample-averaged specific heat $C$. The singularities in $C$ can also be studied by computing the singular part of $\overline{E}_{J}$, as $\overline{E}_{J}$ is just the integral of $C$ over $\sigma$. Thus, one may estimate $\alpha$ from $\overline{E}_{J}$ at $\sigma = \sigma_{\rm c}$~\cite{holm:97} via the scaling
form
\begin{equation}
	\label{eq:E_J_scaling} \overline{E}_{J}(\sigma_{\rm c},L_{\rm eff}) = E_{J,\infty} + b
	L_{\rm eff}^{(\alpha-1)/\nu}(1+b'L_{\rm eff}^{-\omega}),
\end{equation}
where $E_{J,\infty}$, $b$, and $b'$ are non-universal constants.
Since $\alpha^{\rm (MF)} = 0$ and $\nu^{\rm (MF)}=1/2$ above the upper critical dimension as already noted above, it is expected that $(\alpha -1)/\nu = -2$.

\begin{figure}
	\centering
	\includegraphics[width=8.5 cm]{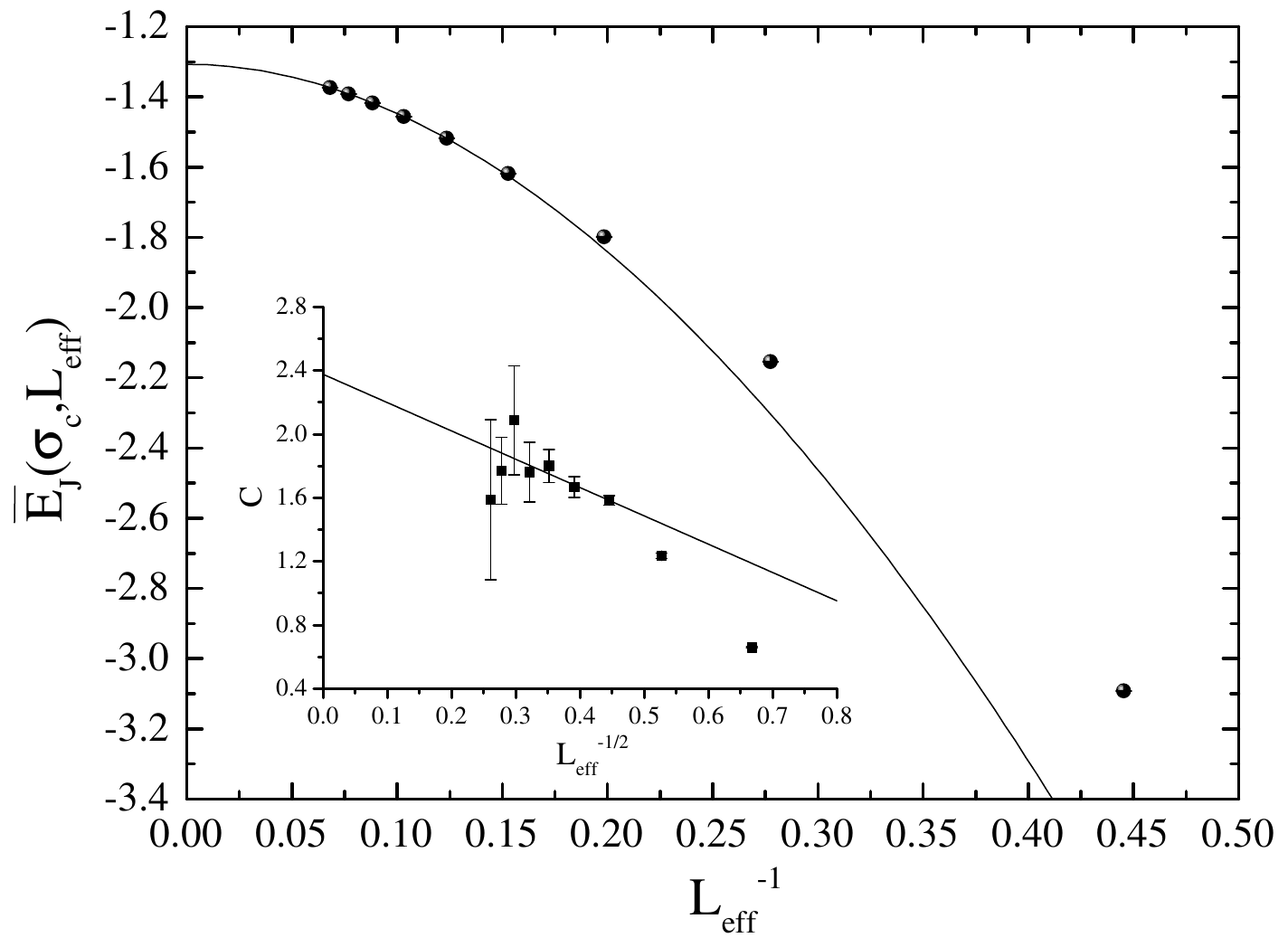}
	\caption{\label{fig:bond_energy_spec_heat} Finite-size scaling behavior of the bond-energy density at the critical random-field strength $\sigma_{\rm c}$ 
		(main panel) and the ``specific-heatlike'' quantity $C$ (inset).}
\end{figure}

Obviously, the use of Eq.~(\ref{eq:E_J_scaling}) for the
application of standard finite-size scaling methods requires an
\emph{a priori} knowledge of the exact value of the
critical random-field strength $\sigma_{\rm c}$~\footnote{An alternative approach based on a three lattice-size variant of the quotients method has been presented in Refs.~\cite{fytas:16b,fytas:17b,fytas:19b}  but is not applicable here due to the limited number of available system sizes.}. Fortunately, we currently have at hand such a high-accuracy estimate of the critical field, see Fig.~\ref{fig:critical_point}. Thus, 
we have performed additional simulations exactly at the critical point
$\sigma_{\rm c}= 9.48391$  for all range of the accessible system sizes using the standard averaging of $10^6$ samples. Data for the bond-energy density are shown in the main panel of 
Fig.~\ref{fig:bond_energy_spec_heat} as a function of $1/L_{\rm eff}$. The solid line is a fair fit ($Q \sim 23\%$) of the
form~(\ref{eq:E_J_scaling}) excluding the smaller system sizes ($L_{\rm min}=5$) while
fixing the exponents $\alpha$, $\nu$, and $\omega$ to their expected values.  

As an additional consistency check we present in the inset of Fig.~\ref{fig:bond_energy_spec_heat}  the scaling behavior of a ``specific-heatlike'' quantity $C$ obtained from the bond-energy density derivative with respect to the random-field strength $\sigma$ at the critical
point $\sigma_{\rm c} = 9.48391$ and using again $10^{6}$ samples. For $C$ the following scaling ansatz is expected
\begin{equation}
	\label{eq:C} C \sim  c_{1}L_{\rm eff}^{\alpha/\nu}(1+c_{2}L_{\rm eff}^{-\omega})  \sim c_{1}+c_{2}'L_{\rm eff}^{-1/2},
\end{equation}
since $\alpha/\nu=0$ at the mean-field level. As it is evident from the plot, the data become rather noisy with increasing system size. This is typical of all derivatives obtained from a fluctuation-dissipation formula. There is no bias, but errors are large because the quantities involved in the fluctuation-dissipation formula are not self-averaging themselves -- see also the discussion Ref.~\cite{fytas:19b}. Therefore we exclude from our fitting attempt the largest system size $L = 10$ where statistical  errors are larger than $30\%$.  The solid line shows a simple linear fit of the form~(\ref{eq:C}) excluding the smaller sizes ($L_{\rm min} = 4$) with an acceptable fitting quality ($Q \sim 89\%$).

\section{Summary}
\label{sec:summary}

We have presented a finite-size scaling analysis of the 7D random-field Ising model with a Gaussian field distribution and periodic boundary conditions. Indeed, above the upper critical dimension the choice of boundary conditions remains crucial~\cite{sola:16}. Ground-state simulations in combination with recent advancements in finite-size scaling and reweighting methods for disordered systems~\cite{fytas:16b} allowed us to provide a high-accuracy confirmation of the mean-field behavior of the model. A major point has been the numerical verification for the use of an effective length-scale $L_{\rm eff} = L^{D / D_{\rm u}}$ (where $D / D_{\rm u} = $ \textgreek{\coppa} in the formulation of Ref.~\cite{berche:22}) in all finite-size scaling relations as has been proposed for the pure Ising ferromagnet~\cite{jones:05,wittmann:14b,kenna:91,kenna:13,berche:12,berche:22} and also the clarification with respect to the corrections-to-scaling exponent $\omega$ in Ising systems above the upper critical dimension. Currently, we are working exactly at $D_{\rm u}$, where characteristic logarithmic scaling violations have been reported~\cite{ahrens:11} but still await for a detailed confirmation. 

\begin{acknowledgments}
We would like to thank Jes\'us Salas for helping us to carry out  numerical checks of the results in Appendix~\ref{appendix}. N.~G. Fytas is grateful to the colleagues in the Department of Theoretical Physics I at Complutense University of Madrid for their warm hospitality, during which part of this work was completed. N.G. Fytas would like to acknowledge the support of EPSRC Grant No. EP/X026116/1. This work was supported in part by Grants No. PID2022-136374NB-C21, PGC2018-094684-B-C21, funded by MCIN/AEI/10.13039/501100011033 by ``ERDF A way of making Europe'' and by the European Union. The research has received financial support from the Simons Foundation (grant No.~454949, G. Parisi).
\end{acknowledgments}

\appendix

\section{Scaling corrections in the large-$N$ limit of the $O(N)$ model for $D>4$}
\label{appendix}

Benefiting from Br\'ezin’s analysis in Ref.~\cite{brezin:82}, we deduce the corrections-to-scaling exponent $\omega$ for the large-$N$ limit of the $O(N)$ model.  

\subsection{General framework}

Let us start by recalling the basic definitions from the original work by Br\'ezin~\cite{brezin:82}. We consider a ferromagnetic system with an $O(N)$-symmetric, nearest-neighbor Hamiltonian on a hypercybic
lattice of linear size $L$
\begin{equation}
  \mathcal{H} = - J N \sum_{\langle\mathbf{x},\mathbf{y}\rangle}
  \vec S_{\mathbf{x}}\cdot \vec S_{\mathbf{y}}\,,\quad
  \vec S_{\mathbf{x}}\cdot \vec S_{\mathbf{x}}=1\,,
\end{equation}
with periodic boundary conditions. From this point and on we shall be using the dimensionless inverse
temperature $\beta = J/T$.

The model greatly simplifies in the limit $N\to\infty$. In the paramagnetic phase,
$\beta\leq\beta_{\mathrm{c}}$, the propagator $[G(\mathbf{r})=\langle \vec S_{\mathbf{x}}\cdot
  \vec S_{\mathbf{x}+\mathbf{r}}\rangle]$ is
\begin{equation}
  G(\mathbf{r})=\frac{1}{\beta}\frac{1}{L^D}
  \sum_{\mathbf{q}}\frac{\mathrm{e}^{\mathrm{i}\mathbf{q}\mathbf{r}}}{m_L^2+\lambda(\mathbf{q})}\,,
\end{equation}
where $\lambda(\mathbf{q})=\sum_{i=1}^D 2(1-\cos q_i)$, $\mathbf{q}=\frac{2\pi}{L}(n_1,n_2,\ldots,n_D)$, $0\leq n_i\leq L-1$, and the mass term $m_L^2$ is the inverse-squared correlation length
$m_L^2=1/\xi_L^2$. One relates $m_L^2$ and $\beta$ through the \emph{gap equation} which simply
codes the constraint $G(\mathbf{r}=\mathbf{0})=1$
\begin{equation}\label{eq:gap-finite-L}
  \beta=\frac{1}{L^D}
  \sum_{\mathbf{q}}\frac{1}{m_L^2+\lambda(\mathbf{q})}\,.
\end{equation}
Note that the dispersion relation $\lambda(\mathbf{q})$ depends crucially on
our choice of the nearest-neighbor lattice interaction. In fact, the only feature shared
by all local-interaction Hamiltonians is
$\lambda(\mathbf{q}\to\mathbf{0})=\mathbf{q}^2+{\cal O}(q_i^4)$.

As it is well-known, the problem becomes much simpler in the thermodynamic limit (where anyway the choice of boundary conditions becomes inconsequential)
\begin{widetext}
\begin{equation}\label{eq:gap-infinite-L}
  G(\mathbf{r})=\frac{1}{\beta}\int_{B.Z.}\frac{\mathrm{d}^D\mathbf{q}}{(2\pi)^D}\,
    \frac{\mathrm{e}^{\mathrm{i}\mathbf{q}\mathbf{r}}}{m_\infty^2+\lambda(\mathbf{q})}\,,\quad
    \beta= \int_{B.Z.}\frac{\mathrm{d}^D\mathbf{q}}{(2\pi)^D}\,
    \frac{1}{m_\infty^2+\lambda(\mathbf{q})}\,,
  \end{equation}
  \end{widetext}
  where $B.Z.$ stands for the first Brillouin zone and $-\pi <
  q_i < \pi$ for $i=1,2,\ldots,D$. Note that the integral in
  Eq.~\eqref{eq:gap-infinite-L} is convergent for $D>2$ even if we plug $m^2_\infty=0$.

The problem we shall be dealing here is the precise connection between
Eqs.~\eqref{eq:gap-finite-L} and~\eqref{eq:gap-infinite-L} as $L$ grows, for
$D>4$. The alert reader will note that this connection cannot be smooth
because of the singular behavior at $m_L^2=0$ and $\mathbf{q}=\mathbf{0}$ (the strong singularity is characteristic of the periodic boundary conditions)
\begin{widetext}
\begin{equation}\label{eq:naif}
  \frac{1}{L^D}
  \sum_{\mathbf{q}}\frac{1}{m_L^2+\lambda(\mathbf{q})}\, =\,
  \frac{1}{L^Dm_L^2}\ +\ L^{2-D}\times \text{(regular term in the limit $m^2_L\to 0$)}\,.
\end{equation}
\end{widetext}
The analysis by Br\'ezin~\cite{brezin:82} puts the above observation in a sound mathematical footing.

\subsection{The (finite) Poisson summation formula}

Let $H(q)$ be a smooth, periodic function $H(q)=H(q+2\pi)$. One starts by recalling the (finite) Poisson summation formula
\begin{equation}\label{eq:Poisson-1}
\frac{1}{L}\sum_{k=0}^{L-1} H(q= 2\pi k/L)=\sum_{n=-\infty}^{\infty}
\int_{-\pi}^{\pi}\frac{\mathrm{d} q}{2\pi}\, H(q)\mathrm{e}^{\mathrm i q n L} \,.
\end{equation}

If the function $H$ depends on a $D$-dimensional argument, $H(\mathbf{q})$, and if
it is periodic (with period $2\pi$) along every one of the $D$ axes in the
$\mathbf{q}$ space, then one can use Eq.~\eqref{eq:Poisson-1} in a nested way
\begin{widetext}
\begin{equation}\label{eq:Poisson-3}
\frac{1}{L^D}\sum_{k_1=0}^{L-1}\ldots \sum_{k_D=0}^{L-1}
H\Big[\mathbf{q}= \frac{2\pi}{L}(k_1,k_2,\ldots,k_D)
\Big])=\sum_{n_1=-\infty}^{\infty} \ldots \sum_{n_D=-\infty}^{\infty}
\int_{B.Z.}\frac{\mathrm{d} \mathbf{q}}{(2\pi)^D}
H(\mathbf{q})\mathrm{e}^{\mathrm i L \mathbf{q}\cdot(n_1,n_2,\ldots,n_D)} \,.
\end{equation}
\end{widetext}
Let us now use the notation $\mathbf{n}=(n_1,n_2,...,n_D)$ and the short hand
$\sum_{\mathbf{n}}$ to refer to the multi-dimensional series in the r.h.s. of
Eq.~\eqref{eq:Poisson-3} [$\sum_{\mathbf{n}}'$ will be the series in which the
term $\mathbf{n}=(0,\ldots,0)$ has been excluded]. Hence, the
gap equation~\eqref{eq:gap-finite-L} can be rewritten as
\begin{widetext}
\begin{equation}\label{eq:gap-finite-L-Poisson}
  \beta=\frac{1}{L^D}
  \sum_{\mathbf{q}}\frac{1}{m_L^2+\lambda(\mathbf{q})}=\, \int_{B.Z.}\frac{\mathrm{d}^D\mathbf{q}}{(2\pi)^D}\,
    \frac{1}{m_L^2+\lambda(\mathbf{q})}\ +\ \sum_{\mathbf{n}}' \int_{B.Z.}\frac{\mathrm{d}^D\mathbf{q}}{(2\pi)^D}\,
    \frac{\mathrm{e}^{\mathrm i L \mathbf{q}\cdot\mathbf{n}}}{m_L^2+\lambda(\mathbf{q})}.
\end{equation}
\end{widetext}
Let us now introduce the notation
\begin{equation}\label{eq:y-def}
  y^2=L^2 m_L^2=\left(\frac{L}{\xi_L}\right)^2\,,
\end{equation}
and analyze the remainder term
\begin{equation}\label{eq:reminder-1}
  R(y,L)\equiv \sum_{\mathbf{n}}' \int_{B.Z.}\frac{\mathrm{d}^D\mathbf{q}}{(2\pi)^D}\,
    \frac{\mathrm{e}^{\mathrm i L
        \mathbf{q}\cdot\mathbf{n}}}{m_L^2+\lambda(\mathbf{q})}\,,\quad
    m_L=y/L\,.
\end{equation}
On the view of Eq.~\eqref{eq:naif}, one may expect for small $m_L$ that
\begin{equation}\label{eq:reminder-2}
  R(y,L)\sim\frac{1}{L^D m_L^2}=\frac{L^{2-D}}{y^2}\,.
\end{equation}
Our analysis is based on the above asymptotic estimate (that we shall now derive). However, because we are interested in corrections to scaling, we shall need to extend this analysis by obtaining as well the next-to-leading
term in Eq.~\eqref{eq:reminder-2}.

Br\'ezin did the following simplification that is only valid at small
$\mathbf{q}$, and which is, fortunately, the regime of interest
\begin{widetext}
\begin{equation}
  \int_{B.Z.}\frac{\mathrm{d}^D\mathbf{q}}{(2\pi)^D}\,
    \frac{\mathrm{e}^{\mathrm i L
        \mathbf{q}\cdot\mathbf{n}}}{m_L^2+\lambda(\mathbf{q})}\approx
  \int_{\mathbb{R}^D}\frac{\mathrm{d}^D\mathbf{q}}{(2\pi)^D}\,
    \frac{\mathrm{e}^{\mathrm i L
        \mathbf{q}\cdot\mathbf{n}}}{m_L^2+
      \mathbf{q}^2}=\int_0^\infty\mathrm{d}t\, \int_{\mathbb{R}^D}\frac{\mathrm{d}^D\mathbf{q}}{(2\pi)^D}\,\mathrm{e}^{-t(m_L^2+\mathbf{q}^2)+ \mathrm i L
        \mathbf{q}\cdot\mathbf{n}}\,.   
\end{equation}
\end{widetext}
In the above expression we used the identity
\begin{equation}\label{eq:truco}
  \frac{1}{A}=\int_0^\infty\mathrm{d}t\, \mathrm{e}^{-t A}\,,
\end{equation}
which allows us to make explicit the integral over $\mathbf{q}$ (which is
now a Gaussian integral)
\begin{equation}\label{eq:Brezin-approx}
  \int_{B.Z.}\frac{\mathrm{d}^D\mathbf{q}}{(2\pi)^D}\,
    \frac{\mathrm{e}^{\mathrm i L
        \mathbf{q}\cdot\mathbf{n}}}{m_L^2+\lambda(\mathbf{q})}\approx
    \frac{L^{2-D}}{(4\pi)^{D/2}} \int_0^\infty\frac{\mathrm{d}t}{t^{D/2}}\,
    \mathrm{e}^{-t y^2 - \frac{\mathbf{n}^2}{4t}}\,,
\end{equation}
where $y$ was defined in Eq.~\eqref{eq:y-def}. Plugging now Br\'ezin's
approximation~\eqref{eq:Brezin-approx} into Eq.~\eqref{eq:reminder-1}, we
obtain
\begin{widetext}
\begin{equation}\label{eq:reminder-3}
  R(y,L)\approx \frac{L^{2-D}}{(4\pi)^{D/2}} \int_0^\infty\frac{\mathrm{d}t}{t^{D/2}}\,
    \mathrm{e}^{-t y^2} g(t)\,,\quad g(t)=
    \sum_{\mathbf{n}}'\mathrm{e}^{-\mathbf{n}^2/(4t)}=\left[\sum_{n=-\infty}^\infty
  \mathrm{e}^{-\frac{n^2}{4t}}\right]^D-1\,.
\end{equation}
\end{widetext}
Note that $g(t)$ behaves for small $t$ as
\begin{equation}
  g(t\to 0) \sim 2D\, \mathrm{e}^{-\frac{1}{4t}}\,,
\end{equation}
hence $g(t)$ regulates the divergence at small $t$ in the integration measure
of Eq.~\eqref{eq:reminder-3} (namely $t^{-{D/2}}$).

We also need a strong command on the behavior of $g(t\to\infty)$. Let $f(x)$
be an (aperiodic) smooth function and $F(q)$ its Fourier transform
\[F(k)=\int_{-\infty}^{\infty}\mathrm{d}\,x\ f(x)\,\mathrm{e}^{-\mathrm{i}2\pi
    k x}\,,\]
then, the Poisson summation formula tells us that
\begin{equation}
  \sum_{n=-\infty}^{\infty}\, f(n) =   \sum_{k=-\infty}^{\infty}\, F(k)\,.
\end{equation}
Using the above identity for $f(x)=\text{exp}(-x^2/4t)$, one obtains
\begin{equation}
  \sum_{n=-\infty}^\infty
  \mathrm{e}^{-\frac{n^2}{4t}}=\sqrt{4\pi t}\left[1+2\sum_{k=1}^\infty
    \mathrm{e}^{-4\pi^2 k^2 t}\right]\,,
\end{equation}
so that one finds for large $t$
\begin{equation}
  \frac{g(t)}{(4\pi t)^{D/2}}\sim 1\ -\ \frac{1}{(4\pi t)^{D/2}}\ +\  2D\, \frac{\mathrm{e}^{-4\pi^2 t}}{\sqrt{4\pi t}}\ldots..
\end{equation}
Plugging this expansion into Eq.~\eqref{eq:reminder-3}, we see that
disregarding the leading term, namely $1$, one would find a convergent
integral even for $y=0$. Hence, we conclude that
\begin{widetext}
\begin{equation}\label{eq:gap-finite-L-Poisson-2}
\beta=\frac{1}{L^D}
  \sum_{\mathbf{q}}\frac{1}{m_L^2+\lambda(\mathbf{q})}=\, \int_{B.Z.}\frac{\mathrm{d}^D\mathbf{q}}{(2\pi)^D}\,
  \frac{1}{m_L^2+\lambda(\mathbf{q})}\quad +\quad R(y = L m_L,L)\,,
\end{equation}
\end{widetext}
with an asymptotic behavior for the remainder term (as $y\to 0$)
\begin{equation}\label{eq:reminder-4}
  R(y,L)=L^{2-D}\left[\frac{1}{y^2} \ + \ {\cal A}\ +\ ...\right]\,,
\end{equation}
where ${\cal A}$ is some constant. The interested reader is  invited to compare
Eqs.~\eqref{eq:gap-finite-L-Poisson-2} and~\eqref{eq:reminder-4} with Eq.~\eqref{eq:naif}.

\subsection{Scaling at the critical point}

Let us consider the gap equation at $\beta=\beta_{\mathrm{c}}$ for an infinite and a finite system
\begin{widetext}
\begin{eqnarray}
    \beta_\mathrm{c}&=& \int_{B.Z.}\frac{\mathrm{d}^D\mathbf{q}}{(2\pi)^D}\,
                        \frac{1}{\lambda(\mathbf{q})}\,,\\
\beta_\mathrm{c}&=& \int_{B.Z.}\frac{\mathrm{d}^D\mathbf{q}}{(2\pi)^D}\,
  \frac{1}{m_L^2+\lambda(\mathbf{q})}\quad +\quad R(y = L m_L,L)\,.  
\end{eqnarray}
\end{widetext}
Taking the difference of the above two equations (and multiplying both sides
of the resulting equation by $L^2$) one obtains
\begin{equation}
  y^2 \int_{B.Z.}\frac{\mathrm{d}^D\mathbf{q}}{(2\pi)^D}\,
  \frac{1}{[m_L^2+\lambda(\mathbf{q})]\lambda(\mathbf{q})}= L^2R(y,L)\,.
\end{equation}
Now, for $D<6$ one gets (${\cal B}$ is some constant)
\begin{widetext}
\begin{equation}\label{eq:correcciones-menos-importantes}
\int_{B.Z.}\frac{\mathrm{d}^D\mathbf{q}}{(2\pi)^D}\,
\frac{1}{[m_L^2+\lambda(\mathbf{q})]\lambda(\mathbf{q})}=
\int_{B.Z.}\frac{\mathrm{d}^D\mathbf{q}}{(2\pi)^D}\,
  \frac{1}{\lambda^2(\mathbf{q})}\ +\ {\cal B} m_L^{D-4} +{\cal O}(m_L^2)\,,
\end{equation}
\end{widetext}
(for $D>6$ the leading correction is of the order of $m_L^2$ and at $D=6$ one expects
something like $m_L^2\log(1/m_L^2)$). Reference~\cite{brezin:82} introduces the notation
\begin{equation}
  \sigma(D)=\int_{B.Z.}\frac{\mathrm{d}^D\mathbf{q}}{(2\pi)^D}\,
  \frac{1}{\lambda^2(\mathbf{q})}\,.
\end{equation}
So, collecting everything and recalling Eq.~\eqref{eq:reminder-4}, we get at the critical point and $D<6$
\begin{widetext}
\begin{equation}\label{eq:todo-en-Tc}
  y^2\left[\sigma(D)+{\cal B} \frac{y^{D-4}}{L^{D-4}} +\ldots\right ]=
    L^2R(y,L)=L^{4-D}\left[\frac{1}{y^2} \ + \ {\cal A}\ +\ ...\right]\,.
\end{equation}
\end{widetext}
Note here that Br\'ezin considered only the case without any corrections to scaling
(i.e. ${\cal A}={\cal B}=0$). In such a case, one gets
\begin{equation}\label{eq:leading-Tc}
  y[\sigma(D)]^{1/4}=L^{\frac{4-D}{4}}\quad\text{or}\quad
  \xi_L(\beta_\mathrm{c})=L^{D/4} [\sigma(D)]^{1/4}\,.
\end{equation}
For the needs of the present work we need to also consider the
corrections-to-scaling terms. Equation~\eqref{eq:todo-en-Tc} can be rewritten as
\begin{equation}
  y[\sigma(D)]^{1/4}=L^{\frac{4-D}{4}}\left[\frac{1+{\cal A}
    y^2+\ldots}{1+\frac{{\cal B}}{\sigma(D)} \frac{y^{D-4}}{L^{D-4}}+\ldots}\right]^{1/4}\,.
\end{equation} 
It is maybe even better to write this in terms of $\xi_L$,
\begin{equation}
  \frac{\xi_L(\beta_\mathrm{c})}{L^{D/4}}= [\sigma(D)]^{1/4} \left[\frac{1+\frac{{\cal B}}{\sigma(D)}\frac{y^{D-4}}{L^{D-4}}+\ldots}{1+{\cal A}
    y^2+\ldots}\right]^{1/4}\,.
\end{equation}
(Note that for $D > 6$, corrections of the order of $y^{D-4}/L^{D-4}$
become corrections of order $y^2/L^2$).

Now, recalling
Eq.~\eqref{eq:leading-Tc}, we see that $y^2\sim L^{(D-4)/2}$. On the other
hand, $(y/L)^{D-4}\sim 1/L^{[(D-4)D]/4}$ (that becomes $1/L^{(D/2)}$ for $D > 6$). Therefore, in the regime $4 < D < 6$ we identify a dominant exponent $\omega_1$ and a subleading one $\omega_2$, as follows
  \begin{equation}
  \label{eq:omega_1}
    \omega_1=\frac{D-4}{2}\,,\quad \omega_2=\frac{(D-4)D}{4}\,.
  \end{equation}
And, of course, one should expect all kind of sub-leading corrections terms, such as
$L^{-2\omega_1}$, $L^{-(\omega_1+\omega_2)}$, etc. 
Relating the result of Eq.~\eqref{eq:omega_1} to the random-field problem (where $D_{\rm u} = 6$ rather than $4$) leads to our main result
  \begin{equation}
    \omega_1=\frac{D-6}{2}.
  \end{equation}
Hence, for the present case of $D = 7$ we obtain $\omega = 1/2$.
  
\bibliography{biblio,biblio_nikos}

\end{document}